\documentclass[10pt, conference]{IEEEtran}

\usepackage{cite}
\usepackage{amsmath,amssymb,amsfonts}
\usepackage{algorithmic}
\usepackage{graphicx}
\usepackage{textcomp}
\usepackage{xcolor}
\def\BibTeX{{\rm B\kern-.05em{\sc i\kern-.025em b}\kern-.08em
    T\kern-.1667em\lower.7ex\hbox{E}\kern-.125emX}}
\usepackage[hidelinks]{hyperref}
\usepackage{cleveref}
\usepackage{siunitx}
\usepackage{tabularx}
\usepackage{balance}

\begin{document}

\title{Work-in-Progress: Real-Time Neural Network Inference on a Custom RISC-V \mbox{Multicore Vector Processor}}

\author{\IEEEauthorblockN{Maximilian Kirschner\textsuperscript{1,2}, Konstantin Dudzik\textsuperscript{1,2}, Jürgen Becker\textsuperscript{1,2}}
\IEEEauthorblockA{\textit{\textsuperscript{1}FZI Research Center for Information Technology}, Karlsruhe, Germany \\
\textit{\textsuperscript{2}Karslruhe Institute of Technology}, Karlsruhe, Germany \\
\{kirschner, dudzik, juergen.becker\}@fzi.de}
}

\maketitle


\begin{abstract}
Neural networks are increasingly used in real-time systems, such as automated driving applications. This requires high-performance hardware with predictable timing behavior.
State-of-the-art real-time hardware is limited in memory and compute resources. On the other hand, modern accelerator systems lack the necessary predictability properties, mainly due to interference in the memory subsystem.

We present a new hardware architecture with an accompanying compiler-based deployment toolchain to close this gap between performance and predictability.
The hardware architecture consists of a multicore vector processor with predictable cores, each with local scratchpad memories.
A central management core facilitates access to shared external memory through a static schedule calculated at compile-time. The presented compiler exploits the fixed data flow of neural networks and WCET estimates of subtasks running on individual cores to compute this schedule.

Through this approach, the WCET estimate of the overall system can be obtained from the subtask WCET estimates, data transfer times, and access times of the shared memory in conjunction with the schedule calculated by the compiler.
\end{abstract}


\section{Introduction}
With the increasing use of machine learning in real-time systems, the performance requirements for the underlying hardware are increasing rapidly. Especially in driver assistance systems and automated driving, using neural networks is essential \cite{survey-autonomous-driving, survey-deep-learning-ad}. At the same time, certification requirements and limited resources necessitate worst-case execution time (WCET) analysis for these systems. Hence, there is a pressing demand for timing-predictable high-performance hardware architectures.

Currently, Graphics Processing Units (GPUs) and specialized accelerator architectures are used to meet the high demand for computing power. However, these heterogeneous systems are optimized for average-case performance and have properties detrimental to timing analysis. Most importantly, they often share a memory with the host CPU to exchange instructions and data, which leads to interference and, consequently, unpredictable behavior.

On the other hand, current real-time architectures that focus on predictability have a significant performance gap to GPUs and Machine Learning (ML) accelerator systems. 

The main contribution of this paper is a hardware architecture and its corresponding compiler toolchain, which addresses the need for the predictable execution of neural networks. The hardware architecture consists of multiple predictable vector processors with local memories and a central management core to access a shared main memory. Loads and stores to the external memory follow a static schedule, pre-calculated at compile-time by the presented tools. 

The remainder of this paper is structured as follows.  \Cref{sec:related-work} gives a brief overview of the related work. Our concept for timing-predictable neural network inference is described in \Cref{sec:concept}. This concept comprises a custom hardware architecture and a compiler-based deployment approach. \Cref{sec:implementation} is concerned with the implementation details of these two components. Finally, \autoref{sec:conclusion} concludes the paper and discusses possible next steps.

\section{Related Work} \label{sec:related-work}
There are fundamentally two prevalent approaches to predictable hardware architectures with high computational performance: multicore processors and vector processors. Our architecture combines both ideas into a multicore processor with a vector extension in each core.
Vector processors follow the Single-Instruction Multiple-Data (SIMD) paradigm, where specialized execution units execute the same instruction on a fixed number of elements in a vector register. Vector processors can be implemented with a relatively simple architecture, where a single stream of instructions passes through a single pipeline. This property is beneficial for predictability. The Vicuna co-processor \cite{vicuna} implements such a vector unit, which was designed with a focus on predictability and can be integrated into different scalar RISC-V processors.
The Vicuna vector co-processor is also an integral component of our implementation.
However, SIMD architectures only solve the problem to a certain extent if the applications provide vector operations that can be mapped to wide vector registers \cite{ara2}.
Multicore processors, in contrast, follow the Multiple-Instruction Multiple-Data (MIMD) paradigm. Multiple cores independently execute different instructions on distinct data. Several predictable multicore architectures \cite{t-crest, parMerasa, interPret} have been proposed. 
When designing a real-time multicore processor, the primary objective is timing-compositionality \cite{timing-compositionality} so that the execution times of tasks running on the individual cores can be estimated and then combined into a total WCET. 
The main challenge for timing-compositional systems is to ensure freedom from interference. 
The existing approaches, therefore, almost all use distributed memory so that each core can execute from its own local memory. TDMA arbitration on the interconnect level is the most common solution for access to shared resources such as off-chip memory or peripherals.
Our concept also relies on local memories, but a schedule calculated at compile time is used for communication with the external memory. As a result, the available bandwidth of the interconnect and the memory can be used more flexibly, allowing for higher maximum throughput.

An alternative approach would be to build application-specific circuits for FPGAs or ASICs. Tools such as hls4ml \cite{hls4ml} or FINN \cite{finn} can be used to build such circuits. However, the development process for application-specific hardware is still very complex and time-consuming. Therefore, in this paper, we present a hardware architecture that supports generic neural networks.

Approaches for the deployment of neural networks in real-time-critical environments can be found in the literature. These works present tools generating predictable code from machine learning models \cite{keras2c, acetone}. The generated code is targeted towards simple, predictable single-core architectures. In \cite{simd-gemm}, a library targeting SIMD architectures is presented. This library implements General Matrix Multiplication (GEMM) routines for predictable execution on COTS processors.
In contrast, in this work, we present a custom multicore architecture with associated deployment tools that extend the deployment to a multicore platform, thereby enabling the inference of more complex models.


\section{Concept for the Timing-Predictable Inference of Neural Networks}\label{sec:concept}
Our platform for the predictable execution of neural networks consists of two parts: a hardware architecture and a corresponding machine-learning compiler toolchain. The hardware offers a high degree of parallelization and guarantees freedom from interference by design, while the compiler is used to deploy neural networks and ensures an efficient execution on the predictable hardware. It partitions the neural network into small subtasks, maps these to the cores, vectorizes the subtasks, and calculates a schedule for main memory accesses.  

\subsection{Hardware Architecture}
\begin{figure}
    \centering
    \includegraphics[width=0.8\linewidth]{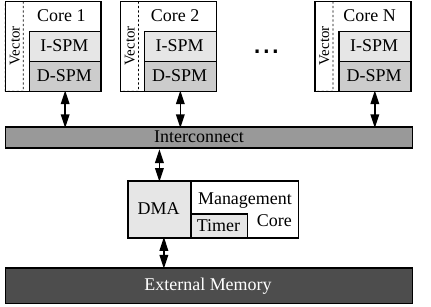}
    \caption{Multicore hardware architecture with local scratchpad memories and vector units in each core}
    \label{fig:hardware-arch}
\end{figure}
The hardware architecture for the predictable acceleration of neural networks is shown in \autoref{fig:hardware-arch}. It consists of $N$ processor cores with an integrated vector unit each, which we refer to as worker cores. A multicore vector processor allows parallelization in two dimensions, distributing computations across the cores and across the vector units in each core. Each core is equipped with two scratchpad memories for instructions and data, respectively. The cores can thus perform computations independently on their local memory.
The architecture also contains a shared external DRAM memory, which is used to store data and instructions. The external memory contains the input data and weights of the neural network and is used to store intermediate results, if necessary.
A central management core orchestrates access to external memory and communication between cores. It copies instructions and data between the scratchpad memories and the external memory, following a static schedule. This schedule is computed by the compiler and specifies communication times for each data movement. A Direct Memory Access (DMA) component with exclusive access to the external memory and the interconnect is used for efficient data transfers. The scratchpad memories are dual-ported so that data can be transferred while the core is executing.

Since the DMA is the only component that can access the external memory and the only one that can initiate transfers on the interconnect, freedom from interference is guaranteed by design. The cores themselves are predictable and operate on their own memory. This allows for separate WCET analyses, which can then be combined with the schedule, which also contains estimates for memory access times and data transfer times, to a total WCET estimation.

\subsection{Compiler-Based Deployment Approach}
Our concept involves MLIR-based \cite{mlir} compiler tools for the deployment of neural networks on the predictable multicore platform. The tools parallelize the model and map it to the cores, vectorize the code for each core, and calculate a schedule for memory access. The input to our toolchain is a trained and quantized ML model. This is then compiled to an optimized RISC-V binary that can be deployed to the external memory.

\begin{figure}
    \centering
    \includegraphics[width=0.85\linewidth]{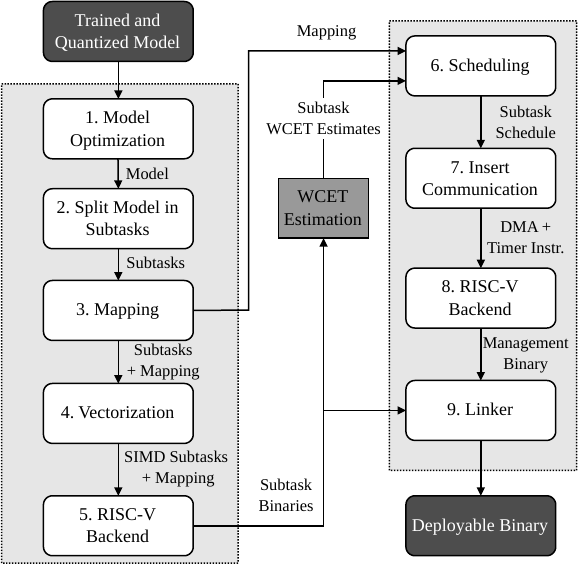}
    \caption{Compilation steps for the deployment of neural networks on the proposed hardware}
    \label{fig:compile-steps}
\end{figure}

\autoref{fig:compile-steps} shows the compilation steps and generated artifacts.
The compiler toolchain consists of two sections as well as an external tool for WCET analysis, which is beyond the scope of this work. The toolchain is split into two workflows, one regarding parallelization, mapping, and vectorization for the worker cores and the other regarding memory access scheduling for the management core. 

In the first step, after loading the model, hardware-independent optimizations, like operator fusion or pruning, are applied to the model.
The model is then split into subtasks that can be executed independently. The size of the subtasks depends on the size of the local memories and the number of cores. In the third step, the subtasks are mapped to the cores. The objective of the mapping is to minimize memory transfers by maximizing data reuse. In the mapping step, the network is traversed in reverse, from the result layer to the input layer, in order to determine dependencies between the subtasks. In a second pass, interdependent calculations are then mapped to the same core to keep as much data as possible in the local memory. Dependencies on subtasks with large amounts of data are prioritized. In the next steps, the programs for the individual cores are vectorized with RISC-V vector instructions and compiled into a RISC-V binary.

An external WCET analysis tool calculates a WCET for each of these artifacts. These estimates and the mapping information are then used as input for the second compiler section.

In the sixth step, a schedule for the subtasks is calculated. The fixed control flow of neural networks without input-dependent branches makes it possible to statically compute such a schedule. The model dictates the order of subtasks. Based on the WCET estimates of the worker core programs, start times for all subtasks are defined. In the seventh step, communication instructions are generated for the management core. Memory transactions are scheduled as early as possible, so that only one transaction takes place at a time. This is achieved using worst-case estimates for the DMA and interconnect transfers as well as the memory access times. If at one point in time several transactions could be executed, these are arranged in the schedule in a round-robin fashion.
The communication and timing methods are then compiled into a RISC-V binary for the management core. In the final step, the binaries for the worker cores and the management binary are linked to decicated address sections to be deployed on the hardware.


\section{Implementation} \label{sec:implementation}
We implement the described hardware architecture and the machine learning compiler for extensive evaluations on an FPGA as part of future work. The following section describes the details of the implementation.

\subsection{Hardware Architecture}
The whole processor is implemented in SystemVerilog, targeting the Xilinx UltraScale+ FPGA architecture. 
The worker cores are based on the open-source RISC-V core Ibex \cite{ibex}, with an attached Vicuna \cite{vicuna} vector co-processor. The Vicuna co-processor implements version 1.0 of the RISC-V vector extension. To be more precise, it implements the \texttt{Zve32x} extension, i.e. the version of the vector extension that is intended for embedded processors. Integer vector elements with 8, 16, and 32 bits are supported. Our implementation is therefore geared towards int8 quantized neural networks. The vector register length of Vicuna can be configured. Our implementation currently uses 512-bit vector registers. 
We have dimensioned the implementation to 16 worker cores with \qty{1}{\mebi\byte} scratchpad memory each, which is divided into dual-ported instruction and data memories. In this configuration, we target the execution of medium-sized convolutional neural networks like ResNet50 or YOLOv5-small. 
The central management core that is responsible for the data transfers into and out of the local memories is a scalar Ibex core that is connected to a modified PULP Data Movement Accelerator (iDMA) \cite{iDMA} and a timer component. The Ibex core is closely coupled to the iDMA and the timer via custom control registers. It configures both components according to the precomputed schedule. The schedule is realized using control and status register instructions from the \texttt{Zicsr} extension.
A TileLink Uncached Lightweight (TL-UL) \cite{tlul} crossbar connects the iDMA with the local memories of the worker cores.  
The iDMA is also connected to the DDR4 memory of the FPGA board via an AXI4 bus.

\subsection{Compiler-Based Deployment Approach} 
We implement the compiler-based concept for the deployment of neural networks on our predictable multicore architecture, using the Multi-Level Intermediate Representation (MLIR) \cite{mlir}. MLIR is a modular framework for compiler development with the aim of reusability. \autoref{tab:mlir-dialects} lists the most important MLIR dialects for the compiler implementation. Open Neural Network Exchange (ONNX) models are the input of our compiler. These are transformed to \texttt{onnx-mlir}. We perform hardware-independent optimizations, tiling, and packing using the \texttt{linalg} dialect on the \texttt{tensor} representation. The tiles that are mapped to the individual cores are vectorized to use the SIMD instructions using the \texttt{vector} dialect. \texttt{affine} operations explicitly represent the memory accesses. Finally, we lower to the \texttt{llvm} dialect and write the binary with the RISC-V backend of LLVM.

For the evaluation we will implement support for the most relevant ONNX layer types, of common CNN architectures, in the compiler. CNNs are common in perception systems in real-time environments and are, therefore, of particular interest. The most compute- and data-intensive layers in CNNs are fully connected and convolution layers. It is important to optimize the mapping, vectorization, and communication schedules for these layers.
If the compiler should later be extended with other layer types, it may be necessary to implement new tiling, mapping and vectorization methods for efficient execution. 


These layers are typically implemented using GEMM routines because they can be efficiently tiled and vectorized. We, therefore, choose to implement these layers in the same way. The subtasks mentioned in the concept, correspond to tiles in GEMM routines, which are mapped to the worker cores. The disadvantage of GEMM-based convolution is data duplication. As access to external memory is expensive in our hardware, the actual duplication of memory is only carried out in the scratchpad.


\begin{table}
    \centering
    \caption{MLIR dialects used for the compiler implementation}
    \begin{tabularx}{\linewidth}{c|X}
       \textbf{Dialect}  &  \textbf{Description}\\ \hline
       onnx-mlir & Implements the Open Neural Network Exchange (ONNX) standard. An ONNX model can be transformed into this dialect. Popular ML frameworks can export models in ONNX format. \\ 
       tensor  & High-level dialect used for tensor representation and tensor modification operations\\
       linalg  & Linear Algebra operations, e.g., matrix multiplication, convolution, or depthwise convolution that operate on buffer or tensor type data. linalg is the entry layer for MLIR code generation and is used for transformations like tiling, fusion, or distribution.\\
       vector  & Represents higher-order vectors, closing the gap between high-dimension tensors and native vectors in hardware. It is used after tiling and mapping to cores, for similar transformations as linalg, but on smaller problems from the perspective of a single compute unit.\\
       memref & Represent references to regions of memory, similar to buffer pointers but with additional shape and type information. It is used for memory planning and concrete data accesses.\\
       affine  & Abstraction for affine operations. Among others, it can represent load and store operations as affine maps, allowing for data reordering without actual modification of data in the memory.\\
       llvm    & MLIR dialect corresponding to the LLVM intermediate representation (IR). This dialect can be translated into LLVM IR for compilation to CPU architectures.\\
    \end{tabularx}
    \label{tab:mlir-dialects}
\end{table}


\section{Conclusion and Outlook} \label{sec:conclusion}
Interference in state-of-the-art accelerator systems used for the execution of neural networks poses a significant challenge for the timing analysis in real-time systems. On the other hand, available timing predictable processors do not provide the necessary compute and memory resources. In this work, we present a custom multicore hardware architecture with local memories and centrally orchestrated memory transactions. For the scheduling of these memory transactions, we propose a compiler-based deployment concept.

Up until now, the main components of the hardware have been implemented, while the compiler-based scheduling and deployment are subject to ongoing work.  
In the next step, we plan to evaluate the hardware architecture in various configurations and assess the trade-offs between the configuration parameters. This includes the number of cores, the length of the vector registers in the cores, and the size of the local scratchpads. This hardware can then serve as a flexible platform for future work on the deployment of machine learning applications in real-time systems.
\balance
\section*{Acknowledgment}
This work was supported by the German Federal Ministry
of Education and Research (BMBF) within the project "MANNHEIM-CeCaS," funding number 16ME0818.

\bibliographystyle{IEEEtran}
\bibliography{literature.bib}
\end{document}